\RequirePackage{fix-cm}
\documentclass[smallextended]{svjour3}       
\smartqed  
\usepackage{graphicx}
\usepackage{siunitx}
\journalname{Journal of Low Temperature Physics}
\begin{document}

\title{Nanomechanical resonators for cryogenic research}


\author{T. Kamppinen \and V. B. Eltsov}


\institute{T. Kamppinen \\
\email{timo.kamppinen@aalto.fi}\\
\at Low Temperature Laboratory, Department of Applied Physics, School of Science, Aalto University, Espoo 02150, Finland
}

\date{Received: 11.07.2018}

\maketitle

\begin{abstract}

Suspended aluminium nanoelectromechanical resonators have been fabricated, and the manufacturing process is described in this work. Device motion is driven and detected with a magnetomotive method. The resonance response has been measured at \SI{4.2}{\kelvin} temperature in vacuum and low pressure $^4$He gas.
At low oscillation amplitudes the resonance response is linear, producing Lorentzian line shapes, and $Q$-values up to 4400 have been achieved. At higher oscillation amplitudes the devices show nonlinear Duffing-like behavior. The devices are found to be extremely sensitive to pressure in $^4$He gas. Such device is a promising tool for studying properties of superfluid helium.

\keywords{NEMS \and Sensors \and $^4$He \and Quantized vortices}
\end{abstract}

\section{Introduction}
\label{intro}

\begin{figure}[h]
	\centering
	\includegraphics[height=4cm]{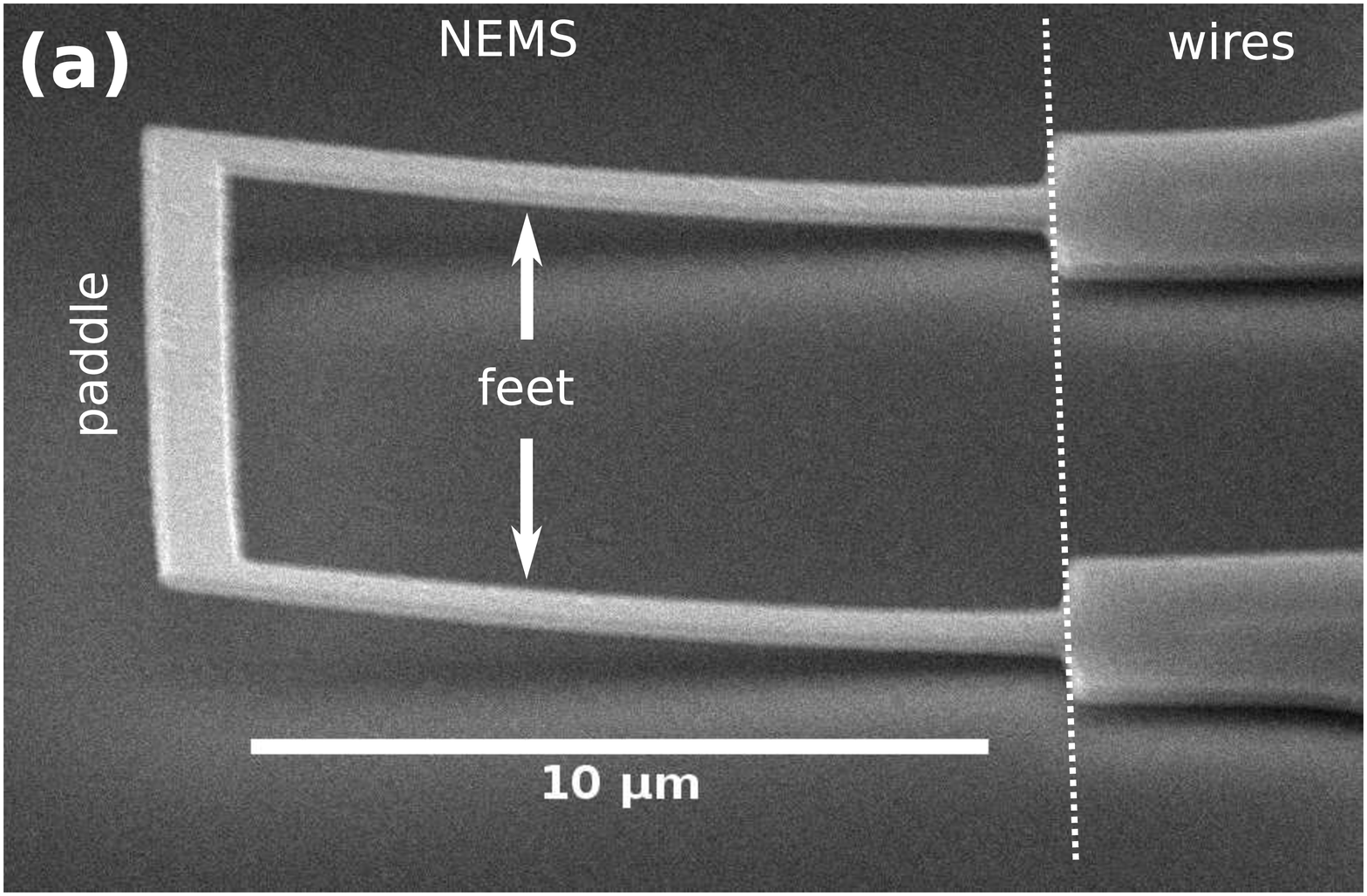}
	\includegraphics[height=4cm]{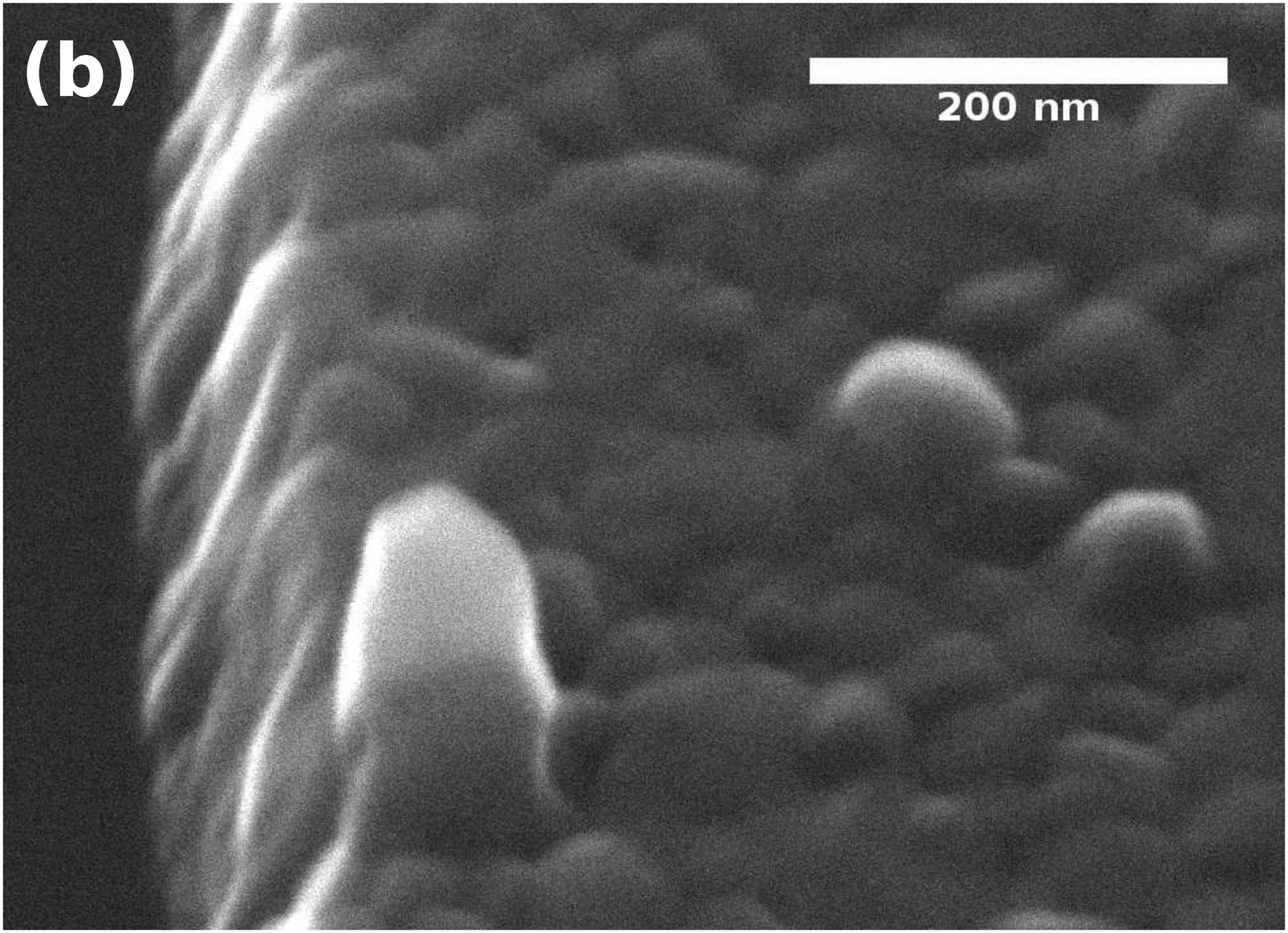}
	\caption{SEM micrographs of one of the samples studied in this work (4NN). {\bf a} The NEMS resonator comprises of two cantilever feet connected by a paddle. Note the upward curvature of the device away from the Si surface. {\bf b} Closeup of the device corner, showing typical grain size ($\sim \SI{50}{\nano \meter}$) and the resulting surface roughness ($\sim \SI{10}{\nano \meter}$) of the evaporated Al film.}
	\label{fig:sample4NN}
\end{figure}

In cryogenic fluids like $^4$He and $^3$He,
immersed oscillating objects such as tuning forks, wires, grids and spheres have proven to be useful and multifunctional tools acting as thermometers, bolometers, pressure gauges, viscometers, as well as generators and detectors of turbulence, cavitation and sound ~\cite{Blazkova2008,Blaauwgeers2007,Pentti2011,Salmela2011}. 
Fluid properties are usually determined from measured changes in mechanical resonance response including resonance frequency, line width, amplitude and certain non-linear effects (for example, nonlinear drag force resulting from turbulence). 
For high sensitivity, a resonator with low mass and spring constant together with a high $Q$-value is required.
Modern micro- and nanofabrication techiques have enabled creation of ultra sensitive probes of the quantum fluids \cite{Bradley2017,Zheng2016,Defoort2016}.
We are pursuing sensitivity to the force resulting from dynamics of a single quantized vortex attached to a mechanical resonator.
This would allow us to study many interesting phenomena, such as Kelvin-wave cascade on a single quantized vortex in $^3$He and $^4$He \cite{Vinen2003,Baggaley2014,Kondaurova2014}; the role of vortex-core-bound fermions in the vortex dynamics in $^3$He-B \cite{Makinen2018,Kopnin1991}; vortex friction due to the chiral anomaly, and the synthetic electromagnetic fields created by vortex motion in Weyl superfluid $^3$He-A  \cite{Bevan1997}.
To reach this goal, we have fabricated suspended aluminium nanoelectromechanical (NEMS) resonators with typical effective mass $\sim\SI{10}{\pico\gram}$, dimensions  $\sim \SI{10}{\micro \meter}$ and rectangular cross section of $\SI{150}{\nano \meter} \times \SI{1.1}{\micro \meter}$ (see Fig. \ref{fig:sample4NN}).
\section{Methods}
\label{sec:methods}
\subsection{Fabrication process}
\label{sec:fabrication}
\begin{figure}[!h]
	\centering
	\includegraphics[width=0.8\textwidth]{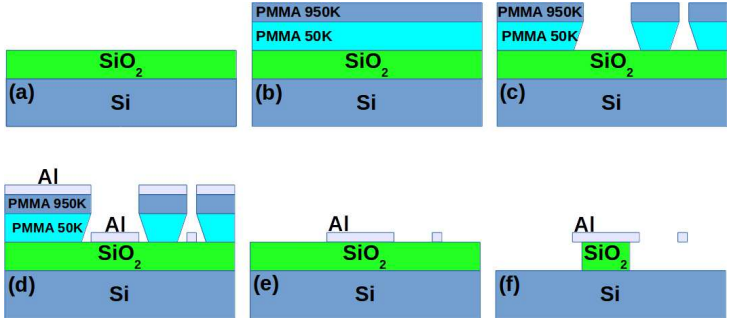}
	\caption{Schematic pictures of the chip after each fabrication step (not to scale): {\bf a}~cleaning, {\bf b}~spin coating, {\bf c}~lithography, {\bf d}~metal deposition, {\bf e}~lift-off, and {\bf f}~release etch. The wide structures, such as bonding pads and wires, remain anchored on top of SiO$_2$ while the nanomechanical resonator is suspended after the isotropic HF vapor release etch.}
	\label{fig:fabsteps}
\end{figure}
The fabrication process of the NEMS devices is presented schematically in Fig.~\ref{fig:fabsteps}.
We start with a $\SI{5}{\milli \meter} \times \SI{5}{\milli \meter}$ high purity (resistivity~$>\SI{100}{\ohm \meter}$) silicon chip with \SI{275}{\nano \meter} of SiO$_2$ on top. First, the chip is cleaned in solvent baths (ethyl pyrrolidinone, acetone and isopropanol) with ultrasound, and a
positive tone PMMA (polymethyl methacrylate, 50k/950k) bilayer is spin coated on top of the chip.
The device pattern is written by electron beam lithography and developed in MIBK:IPA (1:3) solution.
After a brief (\SI{15}{\second}) O$_2$ plasma cleaning,
a \SI{150}{\nano \meter} thick aluminum layer is deposited on the surface of the chip with an electron beam evaporator. To release the vibrating structures, the sacrificial SiO$_2$ layer is etched with isotropic dry HF vapor process. 
The released structure has typically some upward curvature, as seen in Fig.~\ref{fig:sample4NN}. It originates from internal stress formed in the metal layer due to temperature changes during metal deposition. When the device is released the cantilever feet deform and some of the internal stress in the freestanding parts relieves \cite{deformation}.

\subsection{Measurement scheme}
\label{sec:meassetup}
\begin{figure}[b]
\centering
\includegraphics[width=0.7\textwidth]{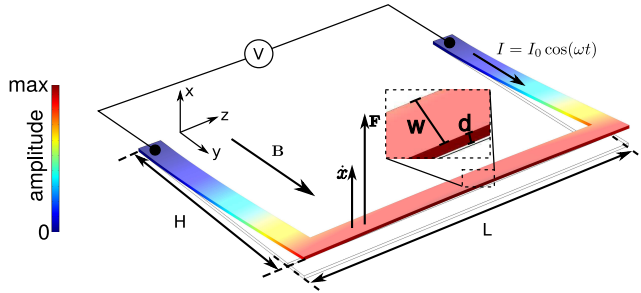}
\caption{Mode shape of the first eigenmode of the sample 4NN, based on Comsol simulation. The magnetomotive drive and detection scheme and geometrical parameters are also depicted. Values of the geometrical parameters are listed in Table \ref{table:resonanceproperties}.}
\label{fig:modeshape}
\end{figure}
\begin{figure}
	\centering
	\includegraphics[width=0.7\textwidth]{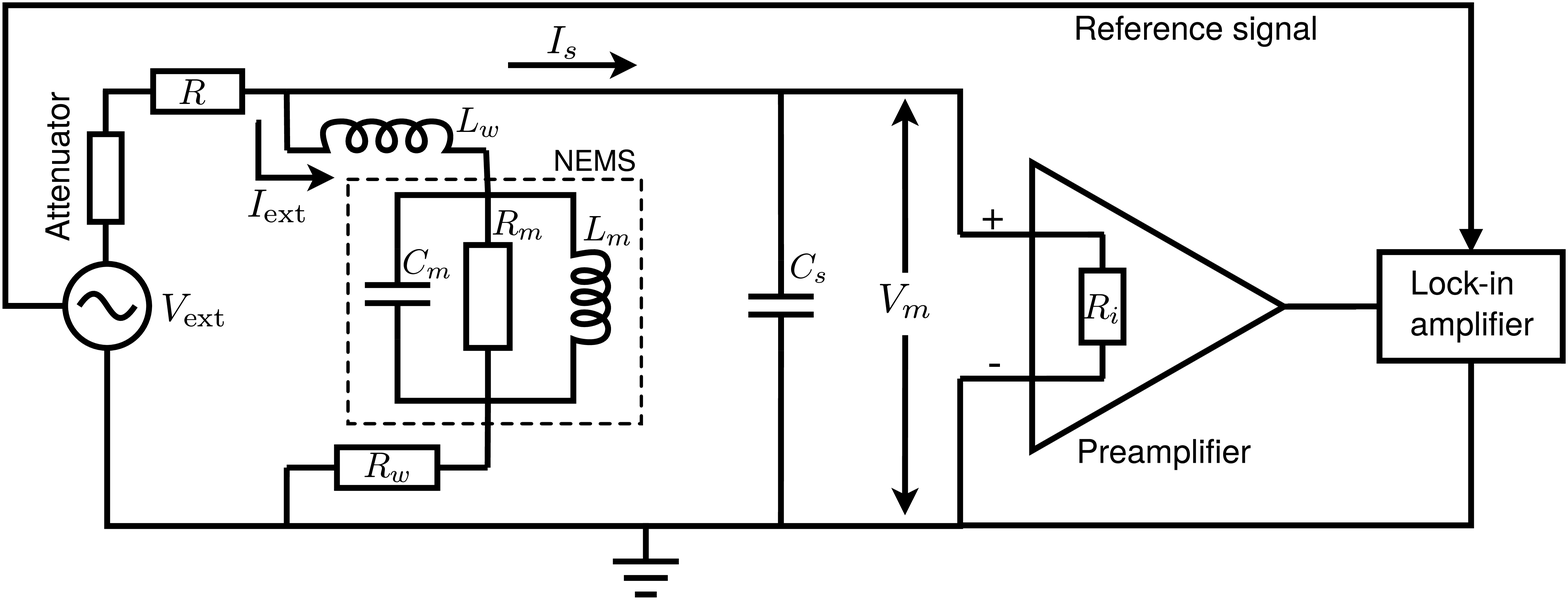}
	\caption{Schematics of the measurement circuitry. The NEMS device is presented with the equivalent RLC resonant circuit in the dashed box. All electronics are at room temperature, while the NEMS device is placed in the middle of a superconducting coil magnet at \SI{4.2}{\kelvin} temperature (not shown in the figure).}
	\label{fig:meassetup}
\end{figure}

\paragraph{Magnetomotive measurement.} The magnetomotive measurement scheme is depicted in Fig. \ref{fig:modeshape}. When a constant magnetic field $B$ is applied in the $y$-direction, perpendicular to the paddle of length $L$, and an AC current $I=I_0 \cos(\omega t)$ is fed through the device, the paddle experiences a Lorentz force $F = I L B$ in the $x$-direction. The force drives the resonator into oscillatory motion at the frequency $\omega$. The first eigenmode of the nanomechanical resonator corresponds to out-of-plane and in-phase oscillation of the two cantilever feet connected by a rigid paddle. In this configuration, the motion of the paddle through the magnetic field generates via electromotive force a voltage $V = \dot{x} L B$ across the paddle, where $\dot{x}$ is the velocity of the paddle. The oscillation amplitudes of velocity $\dot{x}_0$ and displacement $x_0$ are related as $\dot{x}_0 = 2 \pi f x_0$.
\paragraph{Measurement circuit.}
A schematic of the measurement circuitry is presented in Fig. \ref{fig:meassetup}.
The excitation current $I$ is generated by an arbitrary-waveform generator, followed by a \SI{40}{\decibel} attenuator and a \SI{1.2}{\kilo \ohm} resistor connected in series with the resonator.
The voltage over the resonator is amplified with a preamplifier and measured with a lock-in amplifier, which is phase-locked with the generator.
$R_w$, $L_w$ and $C_s$ present the resistance, inductance and stray capacitance of the device, wires and connected devices, and they contribute to a background signal, which is a linear function of frequency for narrow sweeps around the mechanical resonance. This background is subtracted from the measured response in further analysis of the results.
\paragraph{Experimental setup.}
The NEMS devices are wire-bonded to a printed circuit board (PCB) with aluminum wires.
The PCB is attached to a copper plate, which is installed in vacuum chamber of a dipstick setup. The chamber is immersed in liquid $^4$He bath, which provides a stable \SI{4.2}{\kelvin} temperature enabling the use of a superconducting magnet and ensuring low thermal noise. Twisted pairs of copper wires carry the signals between the PCB at \SI{4.2}{\kelvin} and the room temperature end of the dipstick, and standard coaxial cables are used to connect the dipstick to the measurement electronics at room temperature. The magnetic field is produced by a superconducting solenoid.

\subsection{Theorethical background}
\paragraph{Resonance response.} A NEMS device can be treated as damped oscillator driven by an external force $F = F_0 \cos({\omega t})$. At sufficiently small oscillation amplitudes, harmonic approximation is valid. The equation of motion is
\begin{equation} \label{eq:resonatormotion}
\ddot{x} + \gamma \dot{x} + \omega_0^2 x = \frac{F_0}{m} \cos(\omega t),
\end{equation}
where $x$, $\dot{x}$, $\ddot{x}$ are the displacement, velocity and acceleration of the paddle, $\gamma = 2 \pi \Delta f$ is the drag coefficient (in units rad/s), and $\omega_0 = 2 \pi f_0 = \sqrt{k/m}$ is the natural frequency of the mechanical oscillator where $m$ and $k$ are the effective mass and the spring constant. The solution in the frequency domain is given by the Lorentzian functions for absorption and dispersion
\begin{equation} \label{eq:absorption}
\dot{x}_\mathrm{abs}(\omega) =  \frac{\dot{x}_\mathrm{max} \gamma^2 \omega^2}{( \omega^2 - \omega_0^2)^2 + \gamma^2 \omega^2}, \quad \dot{x}_\mathrm{disp} (\omega) =  \frac{\dot{x}_\mathrm{max} \gamma \omega (\omega^2 - \omega_0^2)}{( \omega^2 - \omega_0^2)^2 + \gamma^2 \omega^2},
\end{equation}
respectively.
The maximum amplitude $\dot{x}_\mathrm{max}$ is obtained at the resonance frequency $f_0$, and the full width at half height of the absorption curve is $\Delta f$. The $Q$-value is defined as $Q=f_0/\Delta f$, and it holds for the maximum displacement at resonance that $x_\mathrm{max} = Q F_0 / k$ \cite{Schmid2016}.
At large oscillation amplitudes the response becomes Duffing-like nonlinear \cite{Collin2010} and shows hysteresis depending on the direction of the frequency sweep.
In practice, we measure the voltages $V_\mathrm{abs}=\dot{x}_\mathrm{abs} B L$ and \mbox{$V_\mathrm{disp}=\dot{x}_\mathrm{disp} B L$}, which we convert to velocity or displacement of the beam.

\paragraph{Kinetic damping.}
The force experienced by an oscillating body moving through low pressure gas in the ballistic regime is due to momentum transfer in collisions with individual gas molecules. The moving body experiences kinetic damping \cite{squeeze_film}
\begin{equation} \label{eq:kinetic_damp}
\Delta f_\mathrm{kin} = \left( \frac{8}{\pi^2} \right) \left( \frac{p A^*}{ m \langle v \rangle } \right),
\end{equation}
where $m$ is the effective mass, $ \langle v \rangle $ is the average velocity of the gas molecules, $p$ is the pressure and $ A^* $ is the scattering cross section, which is obtained by weighting the area $A=w(2H+L)$ with the velocity profile over the surface. For the NEMS devices considered here, we have $A^* \approx 0.6 A$.

\paragraph{Squeeze film force.} The vicinity of the Si surface to the oscillating NEMS device results in a squeeze film force, which arises due to compression and decompression of the gas in the narrow gap between the oscillator and the surface. The force has an elastic and a dissipative contribution, and in the ballistic regime the corresponding terms are
\begin{equation} \label{eq:squeeze_film}
\Delta f_{\mathrm{sf}} = \frac{p A^*}{2 \pi m d} \frac{\tau}{1+(\omega \tau)^2} 
\end{equation} 
\begin{equation} \label{eq:squeeze_film_elastic}
\quad k_{\mathrm{sf}} = \frac{p A^*}{d} \frac{(\omega \tau)^2}{1+ (\omega \tau)^2},
\end{equation}
where $d \sim \SI{1}{\micro \meter}$ is the gap between device and Si surface estimated from SEM figures, and
$$\tau = \frac{8A}{\pi^3 \langle v \rangle d } \sim \SI{1E-7}{\second}$$ 
is the diffusion time of $^4$He gas out of the narrow gap at \SI{4.2}{\kelvin} temperature \cite{squeeze_film}.
Comparing the magnitudes of the two additional damping terms in $^4$He gas, we get \mbox{$\Delta f_{\mathrm{sf}}/\Delta f_{\mathrm{kin}} \approx 2.3$}, so squeeze film damping is the dominant loss mechanism of these two. 
The resonance frequency is expected to increase from the vacuum value as $ \delta f = f_0 k_\mathrm{sf} / 2k $ due to the small additional spring constant $k_\mathrm{sf} \ll k$. 
The expected frequency shift for sample 4UV is $ \delta f / p \approx \SI{300}{\hertz / \milli \bar} $ and the increase in line width is expected to be $d(\Delta f) / dp \approx \SI{4}{\kilo \hertz / \milli \bar}$ in $^4$He gas at \SI{4.2}{\kelvin} temperature. 

\section{Results}

\paragraph{Mechanical properties.}
\begin{figure}[!htb]
	\centering
	\includegraphics[width=0.65\textwidth]{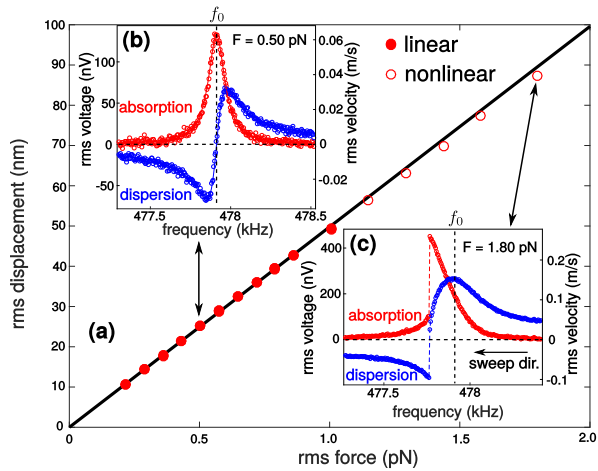}
	\caption{Response of the device 4UN, measured in vacuum at \SI{4.2}{\kelvin} temperature. 
	{\bf a} Displacement amplitude dependence on the excitation force. The black line is a linear fit to the filled circles (linear response regime). 
	{\bf b} An example of linear response with Lorentzian line shapes. Such response is observed up to approximately \SI{50}{\nano \meter} deflections. The resonance properties are obtained from fits to Eq. \ref{eq:absorption} (solid lines). {\bf c} At higher excitation forces the response becomes Duffing-like nonlinear with a negative shift in resonance frequency. The displacement amplitude for panel ({\bf a}) is taken from the absorption peak height, when sweeping frequency in the decreasing direction. The obtained amplitudes deviate from the linear dependence, indicating nonlinear increase in damping at high oscillation amplitudes.}
	\label{fig:extdep}
\end{figure}

The geometrical dimensions and measured properties of the NEMS devices are tabulated in Table \ref{table:resonanceproperties}. The resonance frequency $f_0$ and line width $\Delta f$ are obtained from Lorentzian fits (Eq. \ref{eq:absorption}) to linear resonance spectra, such as shown in Fig. \ref{fig:extdep}. The highest $Q$-value obtained in this work is $Q=4400$ at \SI{4.2}{\kelvin} temperature in vacuum. The measured $Q$-values are in line with typical values reported for NEMS resonators of this size in the literature.
Maximum displacement amplitude as a function of the excitation force for sample 4UN is shown in Fig. \ref{fig:extdep}. 
The effective mass $m$ and the spring constant $k$ are extracted from the slope of the linear dependence using the relations $k=\omega_0^2 m=Q x_{max}/F_0$.
\begin{table}[!h]
\caption{Properties of the three NEMS devices studied in this work. The resonance response is measured at \SI{4.2}{\kelvin} temperature.}
\label{table:resonanceproperties}       
\begin{tabular}{lllll}
\hline\noalign{\smallskip}
Property & 4NN & 4UN & 4UV & description (determined from) \\
\noalign{\smallskip}\hline\noalign{\smallskip}
$L$ (\si{\micro \meter}) & 16.9 & 17.0 & 20.7 & paddle length (SEM micrographs) \\ 
$H$ (\si{\micro \meter}) & 11.0 & 11.1 & 11.0 & feet length  (SEM micrographs) \\ 
$w$  (\si{\micro \meter}) & 1.11 &  1.12 &  1.10 & beam width (SEM micrographs) \\ 
$d$ (\si{\nano\meter})& 150 & 150 & 150 & beam thickness (crystal monitor of evaporator) \\ 
$f_0$  (\si{\kilo\hertz}) & 475.43 & 477.92 & 430.50 & resonance frequency (measurement) \\
$Q$ &  4407 & 3894 & 3115 & Q-value (measurement) \\
$k$  (\si{\newton / \meter}) & 0.08751 & 0.07724 & 0.2262 & effective spring constant (measurement) \\
$m$  (\si{\pico \gram}) & 9.81 &  8.57 & 30.9 & effective mass (measurement) \\
\noalign{\smallskip}\hline
\end{tabular}
\end{table}

\begin{figure}[!t]
	\centering
	\includegraphics[width=\textwidth]{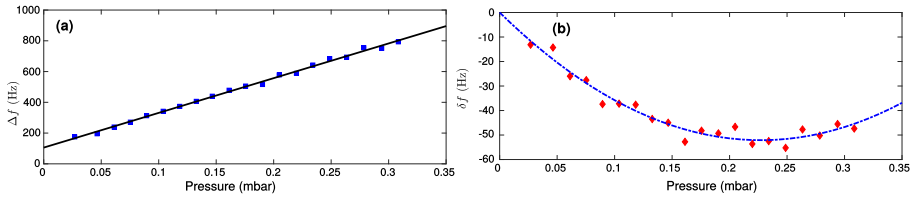}
	\caption{Resonance response of the sample 4UV  as a function of pressure, measured in $^4$He gas at \SI{4.2}{\kelvin} temperature. {\bf a} The resonance line width (blue squares) increases linearly with pressure as expected from theory. The black line is a linear fit. The linear extrapolation to zero pressure gives the vacuum line width for sample 4UV. The slope of the fit is \SI{2257}{\hertz / \milli \bar}, which is 48 \% smaller than predicted by theory (Eqs. \ref{eq:kinetic_damp} and \ref{eq:squeeze_film}).
{\bf b} The measured resonance frequency (red diamonds) shows a minimum around \SI{0.23}{\milli \bar} pressure. The dashed line is a quadratic fit and acts as a quide to the eye. The minimum is a result of two competing effects: increased mass due to deposition of $^4$He atoms on the aluminum surface and the elastic contribution of the squeeze film force.}
	\label{fig:pressuredep}
\end{figure}

\paragraph{Pressure dependence.} We have measured pressure dependence of the resonance response for sample 4UV, and the results are presented in Fig. \ref{fig:pressuredep}. The increase in width with pressure is roughly half the value predicted by theory (Eqs.~\ref{eq:kinetic_damp} and Eq.~\ref{eq:squeeze_film}). The agreement with theory is reasonable, considering that the gap distance $d$ and the value of the diffusion time $\tau$ are not known accurately since the geometry determined from microphotographs at room temperature may change as there are stress induced changes in the device shape as it is cooled down to \SI{4.2}{\kelvin} temperature.
The damping has also been reported to decrease, when the mean free path of gas is close to the gap distance $d$ \cite{Defoort2014}.
From the elastic part of the squeeze film force, we would expect an increase in the resonance frequency.
However, we find a minimum in the frequency around $p=\SI{0.23}{\milli \bar}$, where resonance frequency is shifted by $\delta f_0 = \SI{-52}{\hertz}$. The decrease in resonance frequency can be explained by an additional effective mass $\Delta m_\mathrm{eff} \approx \SI{2.8}{\femto \gram}$ of $^4$He adsorbed on the surface of the oscillator, when pressure is increased. The mass corresponds to one monolayer of $^4$He on the surface \cite{2011JLTP..165...67B}. As the pressure is increased further, the elastic contribution from the squeeze film force (Eq. \ref{eq:squeeze_film_elastic}) takes over and resonance frequency starts to increase.

\section{Conclusions}

Suspended aluminium nanoelectromechanical resonators have been fabricated, and they have been operated in vacuum and $^4$He gas at \SI{4.2}{\kelvin} temperature. $Q$-values up to 4400 have been achieved in vacuum.
The high sensitivity of the devices is demonstrated by large changes in resonance response, as small quantities of $^4$He gas is admitted to the vacuum chamber. Such devices show promise as sensitive probes of the quantum fluids. 




\bibliographystyle{spphys}       
\bibliography{mybib}   

\end{document}